# Thickness dependence of superconductivity in FeSe films

Jia Shi,[1,2] Duy Le,[2] Volodymyr Turkowski,[2]* Naseem Ud Din,[2] Tao Jiang,[2] Qiang Gu,[1]** Talat S. Rahman[2]

[1] *Department of Physics, University of Science and Technology Beijing, Beijing 100083, China*
[2] *Department of Physics, University of Central Florida, FL32816, USA*

* *Corresponding author: E-mail address: <Volodymyr.Turkowski@ucf.edu>*
** *Corresponding author: E-mail address: <qgu@ustb.edu.cn>*

**Abstract**

Thin films of FeSe on substrates have attracted attention because of their unusually high-temperature ($T_c$) superconducting properties whose origins continue to be debated. To disentangle the competing effects of the substrate and interlayer and intralayer processes, we present here results of density functional theory (DFT)-based analysis of the electronic structure of unsupported FeSe films consisting of 1 to 5 layers (1L – 5L). Furthermore, by solving the Bardeen-Schrieffer-Cooper (BCS) equation with spin-wave exchange attraction derived from the Hubbard model, we find the superconducting critical temperature $T_c$ for 1L-5L and bulk FeSe systems in reasonable agreement with experimental data. Our results point to the importance of correlation effects in superconducting properties of single- and multi-layer FeSe films, independently of the role of substrate.

## 1. Introduction

The discovery of superconductivity in $Fe_xSe_{1-x}$ systems has triggered extensive research on these layered materials because of the high transition temperature and several unique properties that come from the presence of the nematic, magnetic and superconducting orders [1-3]. The systems are of interest also because the insights they may provide regarding the role of strong electron correlations in the occurrence of superconductivity. It is important to mention that bulk ("infinite-layer") FeSe undergoes transition to superconducting regime at a notable critical temperature $T_c \sim$ 8K (at ambient pressure) [4]. Besides a not dramatically different value of $T_c$, this material shares a common crystal structure with other Fe-based planar compounds [5-8]. Yet, there is no clear understanding of physics behind superconductivity even in bulk FeSe (i.e., a system with no substrate) because of simultaneous presence of several phase transitions with similar energy order – nematic, magnetic and superconducting. For example, Wang et al. [9] conclude from analysis of their neutron scattering data that spin fluctuations generate both nematicity and superconductivity (this also agrees with theoretical results (see references in [9] and also a recent work [10]). On the other hand, high-temperature (110K) inelastic neutron-scattering measurements [11] demonstrate that there are two types of fluctuations in bulk FeSe - stripe and Néel spin fluctuations – that are seen over a wide energy range. Wang et al. [11] thus conclude that FeSe is "a novel S = 1 nematic quantum-disordered paramagnet interpolating between the Néel and stripe magnetic instabilities". To further clarify the role of nematicity in superconductivity in bulk FeSe, a comparative *ab initio* study of the superconducting properties of the orthorhombic distorted and the higher-energy (not yet discovered experimentally) tetragonal FeSe were performed [12], only to conclude that nematicity is not the driving force for the superconductivity. Another combined experimental-theoretical study of FeSe [13] concluded that high-$T_c$ superconductivity in FeSe based systems could not be explained on the basis of strain driven mechanisms (see a review [14] in which the interplay of nematicity, magnetism and superconductivity in bulk FeSe is discussed). In addition to the complexities mentioned above, Sun et al. [15] (see also Refs. [16,17]) showed that there are indications of a pseudogap in the phase diagram of bulk FeSe, i.e. the fluctuations of the phase of the order parameter, similar to that in cuprate superconductors, might be important in the system (they can be even more important in the 2D case). In short, superconductivity in bulk FeSe continues to vex researchers as it might be the result of several competing transitions.

Recently, a large superconducting gap of about 20 meV and a $T_c$ an order of magnitude higher than that in α-FeSe phase [18-21] have been observed in single-layer FeSe grown on insulating and doped $SrTiO_3$ (STO) substrate through molecular beam epitaxy. Note that despite the simplicity of the Fermi surface of the 1L, the symmetry of the order parameter is still not known. Since scanning tunneling microscopy data for 1L FeSe on STO show scattering between and within the electron pockets, and magnetic impurities suppress superconductivity while non-magnetic ones do not [22], they suggests that the pairing might have s-symmetry. On the other hand, results of a Quantum Monte Carlo study of the 1L FeSe on STO [23] show that phonons play an important role in superconductivity, in agreement with ARPES data [24], and that the pairing can have either s- or d-symmetry depending on the type of spin fluctuations that "glue" the Cooper pairs. The authors also do not rule out importance of the nematic fluctuations in the heavy doped systems that can enhance both s- and d-wave pairings. In general, similar to the bulk case there are evidences of the important role of nematicity [25], fluctuations [26] and stripe order [27] in the superconducting phase in 1L FeSe. For more details, we suggest the following reviews: Ref. [28] for a discussion of possible phonon- and spin-fluctuation exchange mechanisms and Refs. [29,30] for potential topological phases in 1L FeSe/STO and their application in topological-quantum-computation platform. Relevant to this work, scanning tunneling and electron energy loss spectroscopy data [31] strongly supports predominant **electronic pairing mechanism**.

By now, a significant amount of experimental data has been also accumulated for FeSe films on substrates, for which the presence of the substrate also complicates understanding of the mechanism responsible for the superconductivity in these systems. Rather unexpectedly, it was found [32] that contrary to the case of the monolayer, bilayer FeSe film on STO is an insulator, most probably as a result of the strongly-reduced doping efficiency in the bottom FeSe layer as compared to the 1L FeSe–STO interface (for a review of the superconducting properties of films, see Ref. [33]). Furthermore, it was found from transport measurements [34] that 3L and 5L FeSe films on STO show the same $T_c = 40K$ as the 1L, while the normal resistivity decreases with increasing thickness. These results also suggest that the superconductivity emanates at the interface and possibly in the first FeSe layer. It has also been argued [34] that the charge transfer from the doped substrate is crucial for high $T_c$.

To disentangle the effect of the substrate in influencing the superconductivity of films, it is important to analyze the superconducting properties of isolated (free) systems. There are also other reasons for such a study. For example, it was demonstrated in Ref. [35] that electrochemically etched ultrathin FeSe transistor on MgO demonstrates $T_c = 40$ K, i.e. shows that external field generates high $T_c$ in a thin film, regardless of the substrate (with critical thickness of 10 layers for high-$T_c$ superconductivity), i.e. the superconductivity must be generated in the FeSe subsystem. Another example can be found in Ref. [36] in which a thin FeSe flake with $T_c$ less than 10K was doped by a gate voltage to display a high temperature superconductivity onset at 48 K.

To better understand the properties of films, it is natural to establish at first the differences in the physical properties of the FeSe bulk and film systems. It is already known that, contrary to bulk FeSe, in 1L FeSe (system with the highest $T_c$ among the class of iron-based superconductors [37-39]) the Fermi surface has no hole pocket at the Brillouin zone center, and that the Fermi surface topology varies dramatically with increasing thickness of the films [24,40]. As for the origin of the superconductivity in FeSe films, including the pairing mechanism, it is a hotly debated topic [41-46]. It is generally accepted that phonon-mediated pairing can be excluded since electron-phonon coupling in these systems is too weak to overcome the Coulomb repulsion and obtain a high $T_c$ [47,48] (also, no isotope effect has been detected in these systems [49,50]). Thus, it has been proposed that strong electron correlations play an important role in these structures [51-53]. In fact, neutron inelastic-scattering measurements demonstrated that in FeSe magnetic scattering in momentum space is very close to the wave vector $(\pi, \frac{\pi}{2})$, which means that magnetic order in FeSe might differ from the collinear AFM. In the non-superconducting phase, FeSe does exhibit spin fluctuations at low temperatures [54]. Furthermore, high-pressure experiments confirmed the presence of spin density wave (SDW) in bulk FeSe and thin FeSe films, and it was established that magnetic fluctuations (SDWs) are pivotal for the stabilization of high-Tc superconductivity in the bulk and flakes/thin films of FeSe on SrTiO$_3$ under pressure [40,55]. Thus, we included the spin degree in the calculations, with spin arrangements set as checkerboard AFM order obtained theoretically by Cao et al. [56] that allowed to explain insulator-superconductor transition in 1L FeSe. (Our finding indicates that the total energy of the AFM state is lower than the nonmagnetic (NM) state, and magnetic moments (Table I below) are consistent with the theoretical result for the magnetic moment 1.7 for both bulk and monolayer FeSe in the AFM checkerboard phase case [56]). For other theoretical studies of the spin-fluctuation mechanism of the superconductivity in FeSe, we refer the reader to works [57] (bulk) and [2,58,59] (films). Very relevant to the work here, an important study [60] of the possibility of spin-wave- (i.e., electron-correlation-) induced superconductivity in several FeSe systems – isolated monolayer, monolayer on STO and bulk – was performed by using a combined quasi-particle self-consistent GW (QSGW) and DMFT approach. It was found that in these systems with five Fe d-bands near the Fermi surface correlations are orbital-selective, i.e., "Hundness" (described by the exchange parameter J) might play an important role in the properties of the materials, including the value of $T_c$. It was also found that $d_{xy}$ is the orbital with strongest correlations, in agreement with experimental data [61], and that the results strongly depend on the value of J. In bulk FeSe, both DFT, QSGW and DMFT predict that the $d_{xy}$ orbital crosses the Fermi level (in disagreement with ARPES data which shows that the $d_{xy}$ band is 17meV below the Fermi surface [60]). Since in the isolated 1L FeSe the $d_{xy}$ band is even further from the Fermi level (below by 300meV, the energy of magnetic excitations) [60], spin-fluctuations and hence superconductivity is suppressed. On the other hand, as the authors claim [60], the STO substrate changes the situation: the substrate leads to a larger number of carriers and moves the $d_{xy}$ closer to the Fermi energy (still separated by 50-100meV, in agreement with ARPES data). Thus, on the basis of the spin-wave exchange mechanism it was concluded [60] that the $d_{xy}$ band is mostly responsible for the enhancement of the superconducting temperature from 9K in the bulk to ~45K in the 1L FeSe/STO (this calculated $T_c$ is lower than the experimental value 75 K). To summarize, the authors found that the main reasons for high $T_c$ in 1L FeSe on STO substrate are J (larger Hund's correlations correspond to a lesser electron screening) and (substrate-induced) proximity of the $d_{xy}$ band to the Fermi energy. In other words, combined effect of electron correlations and of the substrate are responsible for high-$T_c$ in the system.

In this work, to quantify the role of correlations in the films we perform first-principles calculations of the electronic structure of isolated 1L - 5L FeSe systems, the results of which are then used to analyze superconducting properties of the systems, assuming an antiferromagnetic (AFM) spin wave-mediated pairing corresponding to the proposed spin-fluctuation within (π, π) nesting vector [57,62]. As we show, based on a rather simple model, that *allowed us to perform a comparative analysis of the properties of the complex multilayer systems*, we are able to obtain superconducting critical temperatures that are not too different from what has been observed for the corresponding systems on the substrate, suggesting that spin-mediated pairing can be dominant in both 1L and multi-layer systems on STO substrate.

## 2. Methodology

Our first-principles calculations were carried out using the Vienna *ab initio* simulation package (VASP) [63] within the framework of density functional theory+U (DFT+U). Apart from the fact that we need to use DFT+U instead of DFT for systems in which electron correlations are important, DFT produces unobservable hole pockets in the Fermi surface which is contrary to experimental observations (see below). The projected augmented wave (PAW) pseudopotential was applied to describe the effect of ions, and the exchange/correlation effects were included by using the generalized gradient approximations (GGA) in the form of the Perdew-Burke-Ernzerhof (PBE) potential [64,65]. The one-electron wave functions were expanded in a plane-wave basis with a cutoff energy of 500 eV. The Brillouin-zone was built by using (13×13×1) k-point sampling in the Monkhorst-Paxton mesh. The self-consistent convergence to the tolerance was set at $1\times10^{-6}$ eV/Å, and the maximal force on atoms for the total energy convergence was chosen to be 0.001 eV/Å.

α-FeSe occurs naturally in a tetragonal phase and its symmetry is determined by the P4/nmm space group. Our model supercell consists of n=1-5 layers of FeSe, constructed by using the optimized lattice constant of 3.76 Å and a vacuum slab of 15 Å that separates periodical images along the direction normal to the FeSe film. The model supercells of the studied five FeSe films of different thickness are shown schematically in Fig. 1.

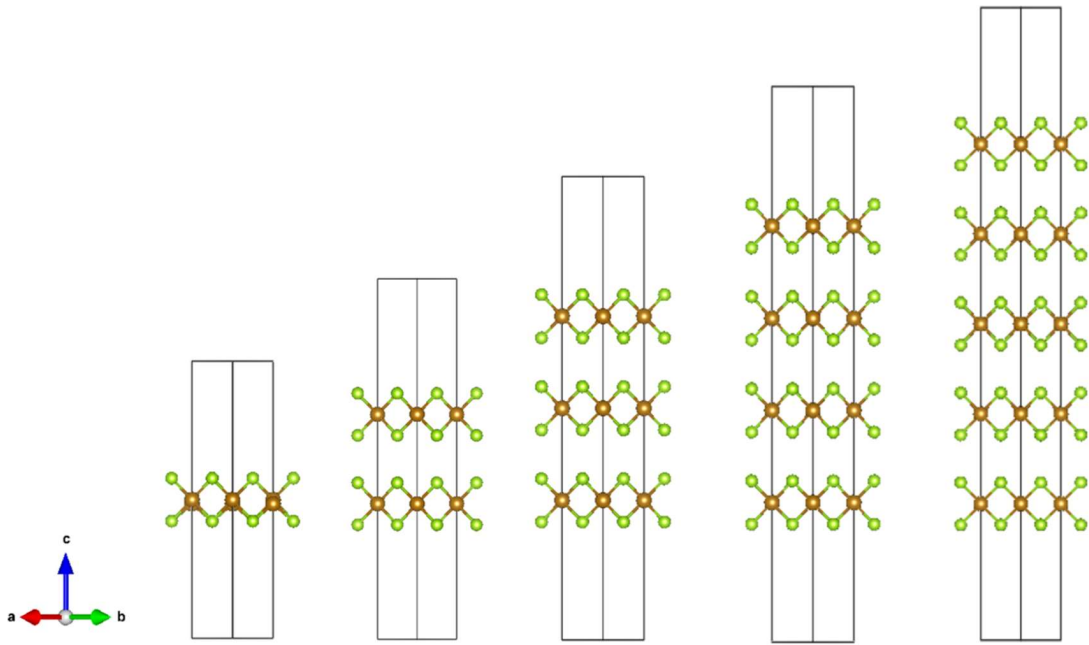

**Fig. 1.** The model supercells of the FeSe films with thickness ranging from 1L (left) to 5L (right). The 15 Å vacuum "sticks" along the C axis are also shown. The green circles refer to Se and brown

- to Fe atoms.

Our calculations show that the total energy of the checkerboard AFM state of the systems is lower than the nonmagnetic (NM) one. Results of our calculations for the magnitude of the magnetic moment for different systems at U=0 are summarized in Table I (and at finite U's - in Supplementary Information (SI), Section 6). The results are consistent with the theoretical result for the magnitude of the magnetic moment 1.7 for both bulk and monolayer FeSe in the same phase obtained by Cao et al. [56].

**Table I.** Modulus of the magnetic moment per layer in the AFM checkerboard phase in FeSe systems.

| Systems | 1L ($\mu_B$) | 2L ($\mu_B$) | 3L ($\mu_B$) | 4L ($\mu_B$) | 5L ($\mu_B$) | Bulk ($\mu_B$) |
|---|---|---|---|---|---|---|
| Magnetic moment | 1.792 | 1.787 | 1.786 | 1.762 | 1.780 | 1.709 |

In this work, we consider spin fluctuation-mediated scenario of superconducting pairing generated by the appropriate part of the Hamiltonian that can be derived by using the Hubbard model (details of the formalism can be found in Ref.[66], and for similar approaches for FeSe and iron superconductors see a recent review [67]). In the case of the AFM spin wave-exchange interaction the total superconducting Hamiltonian has the following form (the free electron part plus the interaction part that also defines spin fluctuations):

$$H = \sum_{l,\vec{k},s} \varepsilon^l(\vec{k}) c^+_{l\vec{k}s} c_{l\vec{k}s}$$

$$-\frac{1}{2N^2} \sum_{\substack{l,m\vec{k},\vec{k}',\vec{q} \\ s_1-s_4}} v(\vec{q}) \vec{\sigma}_{s_1 s_2} \vec{\sigma}_{s_3 s_4} c^+_{l,\vec{k}+\vec{q}\, s_1} c_{l,\vec{k} s_2} c^+_{m,\vec{k}'-\vec{q} s_3} c_{m,\vec{k}' s_4}, \quad (1)$$

where l, $\vec{k}$, and s are the orbital (we use five d-orbitals in our calculations), momentum and spin indices, $\vec{\sigma}_{s_1 s_2}$ are the Pauli matrices,

$$v(\vec{q}) = U + U^2 \chi(\vec{q}) \quad (2)$$

is the electron-electron interaction, U is the local Coulomb repulsion, and N is the number of sites in the lattice.

Since in DFT+U the correlation effects are already included at the quasi-particle level, in our analysis in Eq. (2) we use the one-loop approximation for calculating the susceptibility:

$$\chi(\vec{q}) = \chi_0(\vec{q}), \qquad (3)$$

where

$$\chi_0(\vec{q}) = -i \sum_{l, s_1 - s_4} \int \frac{d\omega'}{2\pi} \int \frac{d^2 q'}{(2\pi)^2} \sigma^z_{s_1 s_2} G_{l, s_2 s_3}(\omega', \vec{q}') \sigma^z_{s_3 s_4} G_{l, s_4 s_1}(\omega + \omega', \vec{q} + \vec{q}') \qquad (4)$$

is the "free-electron" susceptibility

In Eq. (4), $G_{l s_1 s_2}(\omega, \vec{q})$ is the Fourier-transformed single-electron retarded Green's function

$$G_{l s_1 s_2}(t, \vec{r}) = -i\theta(t)\langle c_{l s_1}(t, \vec{r}) c^+_{l s_2}(0,0)\rangle = -i\theta(t)\delta_{s_1 s_2}\langle c_{l s_1}(t, \vec{r}) c^+_{l s_1}(0,0)\rangle \qquad (5)$$

that has the following form in frequency-momentum representation:

$$G_{l s_1 s_2}(\omega, \vec{q}) = \delta_{s_1 s_2} \frac{1}{\omega - \varepsilon^l(\vec{q}) + i\delta}, \qquad (6)$$

where $\varepsilon^l(\vec{q})$ is the DFT+U dispersion of the $l$th band. To take into account multi-orbital effects, we calculate susceptibility with contribution of all d-bands.

After the electron susceptibility $\chi(\vec{q})$ was calculated, we approximated it by a function to ensure that it has maximum at $\vec{q} \equiv \vec{Q} = \left(\frac{\pi}{a}, \frac{\pi}{a}\right)$ corresponding to the AFM spin-wave exchange interaction:

$$\chi(\vec{q}) \approx \bar{\chi}_0 [1 - b(\cos q_x + \cos q_y)], \qquad (7)$$

where parameter $b > 0$ for AFM spin arrangement (in the ferromagnetic case, $b < 0$) and $\bar{\chi}_0$ is a momentum-independent parameter. Parameters $b$ and $\bar{\chi}_0$ were obtained by fitting the susceptibility (7) to the exact function (3).

Next, we used the fact that the electron-electron attractive interaction can be factorized:

$$V_{\vec{k}\vec{k}'} = \sum_n V_n \gamma_n(\vec{k}) \gamma_n(\vec{k}'), \qquad (8)$$

where $\gamma_n(\vec{k})$ is the symmetry function in channel n.

In particular, in the attractive singlet-pairing extended s-wave channel, one has

$$\gamma_{s(e)}(\vec{k}) = \cos k_x + \cos k_y, \quad V_{s(e)} = -\frac{3}{4}U^2\bar{\chi}_0 b \tag{9}$$

and in the also attractive d-wave channel, considered in this paper, the corresponding functions are

$$\gamma_d(\vec{k}) = \cos k_x - \cos k_y, \quad V_d = -\frac{3}{4}U^2\bar{\chi}_0 b \tag{10}$$

(the interaction strengths $V_{s(e)}$ are equal in both channels). To approximate exact susceptibility (see SI, Section 7) by Eq. (7), we put $b = 0.5$ and $\bar{\chi}_0 = (\chi(\pi,\pi) + \chi(0,0))/2$. Which gives a susceptibility with maximum at $\vec{q} = (\pi,\pi)$ and minimum at $\vec{q} = (0,0) = (2\pi, 2\pi)$, which is a reasonable approximation for susceptibility for the AFM spin-exchange scenario of superconductivity (indeed, as it follows from SI, Section 7, exact susceptibility has a sharp maximum at $\vec{q} = (\pi,\pi)$).

In this case, for the superconducting gap one has the following function:

$$\Delta_n(\vec{k}) = \Delta_{0n}\gamma_n(\vec{k}), \tag{11}$$

where $\Delta_{0n}$ is a channel-dependent parameter. The corresponding channel-dependent BCS equation for the critical temperature is

$$1 = V_n \int \frac{d^2k}{(2\pi)^2} \gamma_n^2(\vec{k}) \frac{\tanh\frac{\varepsilon(\vec{k})}{2T_c}}{\varepsilon(\vec{k})}, \tag{12}$$

where $\varepsilon(\vec{k})$ is dispersion of the conduction band. Equation (12) is the equation we solved to find the superconducting critical temperature of the layered systems at different values of U.

## 3. Results and discussions

### A. Electronic structure

The calculations of the electronic properties are performed to yield equilibrium (relaxed) atomic positions with the fixed structural volume of the AFM configuration. As one can see from Fig. 2 and SI Section 1 (and for bulk – SI, Section 3), at first sight the dependence of density of states (DOS) of the films with dominant Fe-3d contribution on their thickness is not pronounced, though as we show below the difference in the DOS near the Fermi energy plays crucial role. As it follows from Fig. 2(a) and Fig. S9(a) in SI, DFT results for the band structure of bulk FeSe are in agreement with those from calculations previously reported [62], suggesting the validity of our approach and results. Using the computational setting of the bulk system, we find that the band structure of the 5L system is quite similar to that of bulk FeSe. Also, for the 1L case, our results are in agreement with those obtained in prior DFT calculations [68] and experimental data (see Supplementary Information in Ref. [38]) which reports signatures of flat bands. Moreover, the flat bands obtained in DFT calculation [68] survive also in our DFT+U results (Fig. 2 and Figs. S2-S5).

The corresponding Fermi surfaces in the 1L case are shown in Figure 3 (for other systems, see SI, Sections 2,4). Importantly, DFT+U calculations give only electron pockets around the M points and hole pockets are absent (contrary to the DFT case).

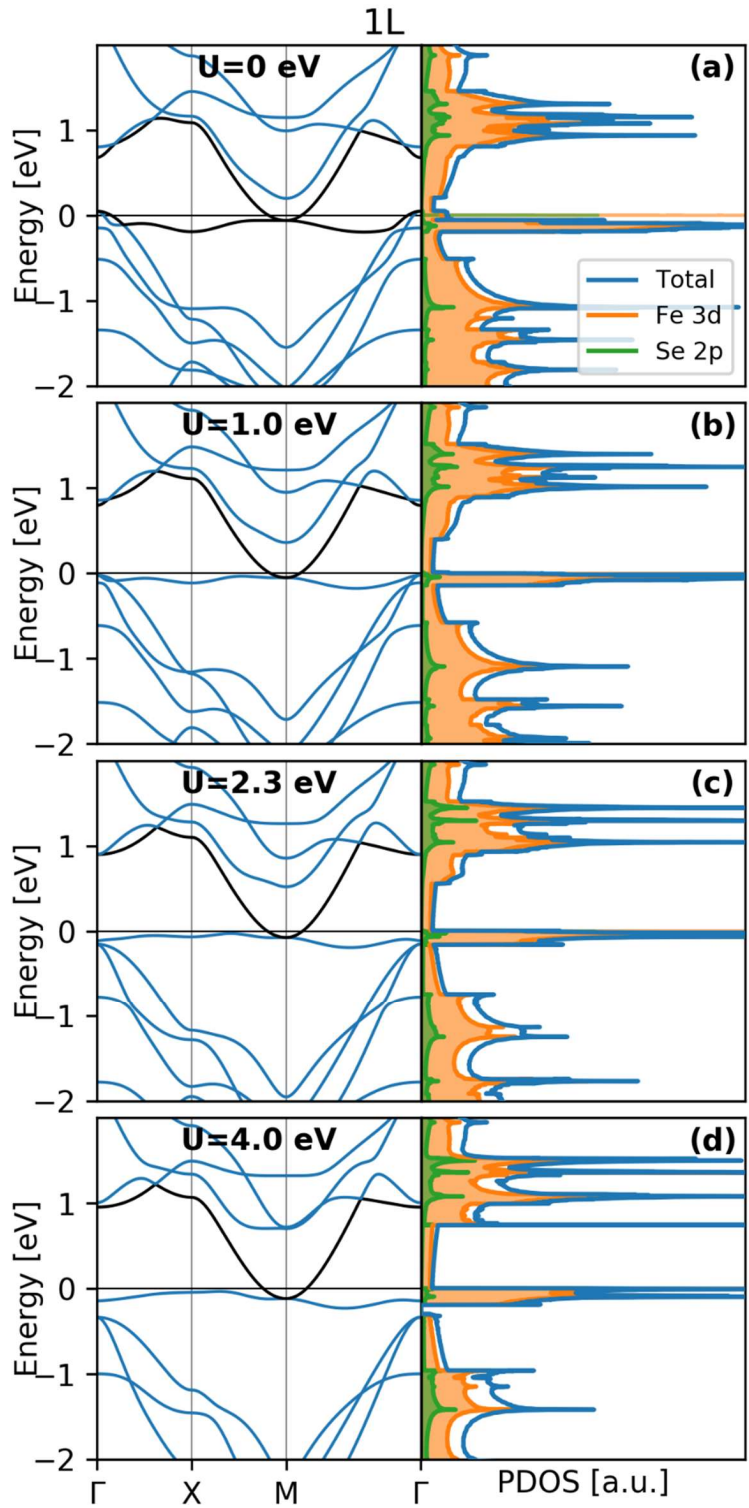

**Fig. 2.** The total and projected Fe (3d) and Se (2p) DOS for the 1L FeSe system at different values of U.

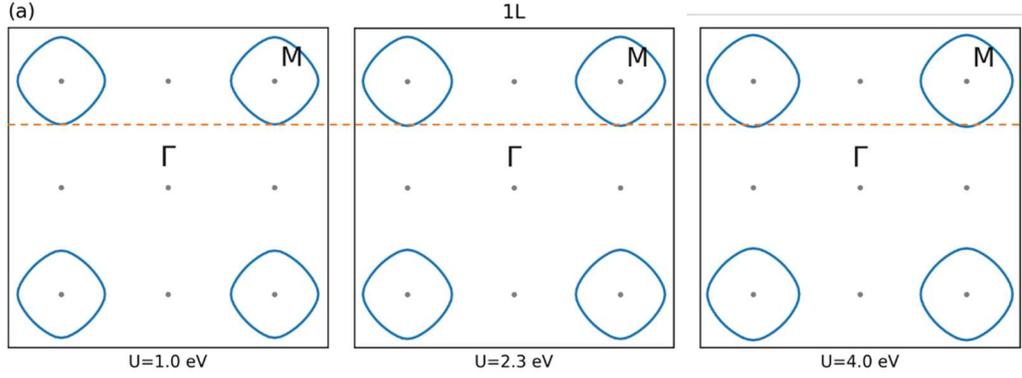

**Fig. 3.** The Fermi surface of the 1L FeSe system at different values of U.

According to BCS [69] and other theories, the DOS at the Fermi level $N(0)$ is the key quantity that defines a possibility to have the superconducting condensate (the critical temperature and the gap in the BCS theory are proportional to $\exp(-1/2\pi\lambda N(0))$, where $\lambda$ is the coupling constant; see also Eq.(12), where one can express momentum integration in terms of the energy integration with corresponding DOS). Thus, it is important for opening superconducting gap whether there is high DOS around the Fermi level. It is clear from Fig. 2 that $N(0)$, with main contribution coming from the Fe(3d) states, is rather high for all 1L - 5L systems. Thus, the dominating in the DOS 3d states of Fe, and spin waves they "generate", may play the key role in forming the superconducting condensate in the FeSe films.

## B. Superconducting critical temperature

Using equation (12), we calculated the transition temperature in 1L-5L FeSe films at different values of U. The results presented in Fig. 4 show that $T_c$ grows with increasing U and decreasing number of layers in the film. The results for critical temperature for 1L FeSe are in a rather good agreement with the experimental Tc ~ 65K for 1L FeSe grown on the STO substrate [18] for all values of U. Also, the results for the bulk FeSe at different U's are not too different from the experimental critical temperature (~ 8K), thus supporting meaningfulness of the approximations that we have used in this work.

One can understand the reason for difference in critical temperature for systems with different number of layers through their normalized (per layer) total DOS (shown in Fig.5, and for bulk in Fig. 6, see also SI, Section 5). Namely, the difference can be attributed to "melting" of the peak at ~-0.07eV and to shift of the "main" peak at ~-0.03eV as the number of layers increases (left Fig. 5). The above can semi-quantitatively explain similarity of the results within two groups (1L-3L and 4L-5L) of the considered systems.

It must be noted that we have obtained rather high critical temperatures, even though the $d_{xy}$ orbital (claimed in several works mentioned above to be the dominant one) is not close to the Fermi level (similar DFT results for the spectrum were obtained, e.g., in Ref. [70]). The main reason for this is collective effect of several spin-wave modes generated by different Fe d-orbitals and much closer in energy average position of the d-bands to the Fermi energy (~ 30meV) as compared to the QSGW results (~150meV) [80]. Another important result is the lack of strong sensitivity of $T_c$ to the Coulomb repulsion parameter, as long as it lies in the range of "realistic" U's~1-4eV, which suggests reliability of the results. We expect that substrate might modify more significantly the average-per-layer electronic DOS of 1L and 2L systems and not the spin-wave contribution to $T_c$, at least in the 3L-5L systems. On the other hand, as results of our comparative analysis show the spin-wave contribution is basically of the same order of magnitude for all systems, even if they are put on a substrate.

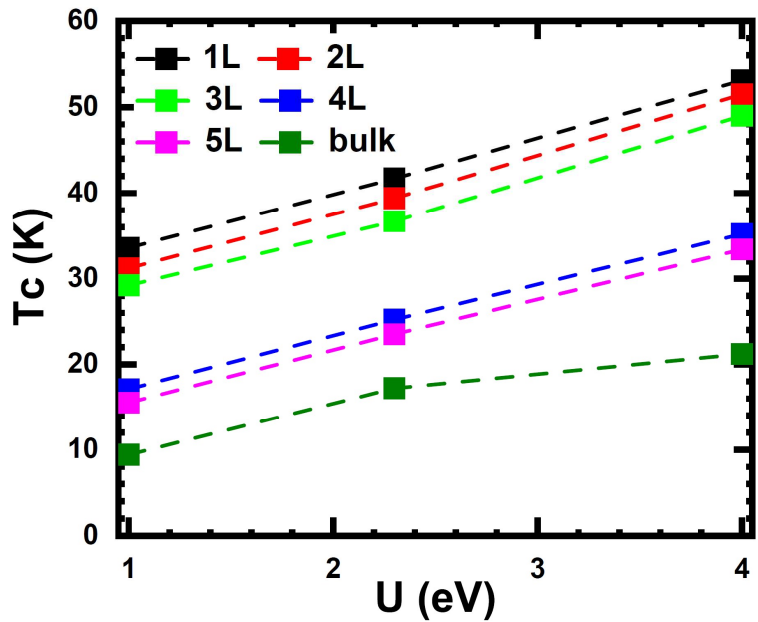

**Fig. 4.** Dependence of the critical temperature on U for different FeSe systems.

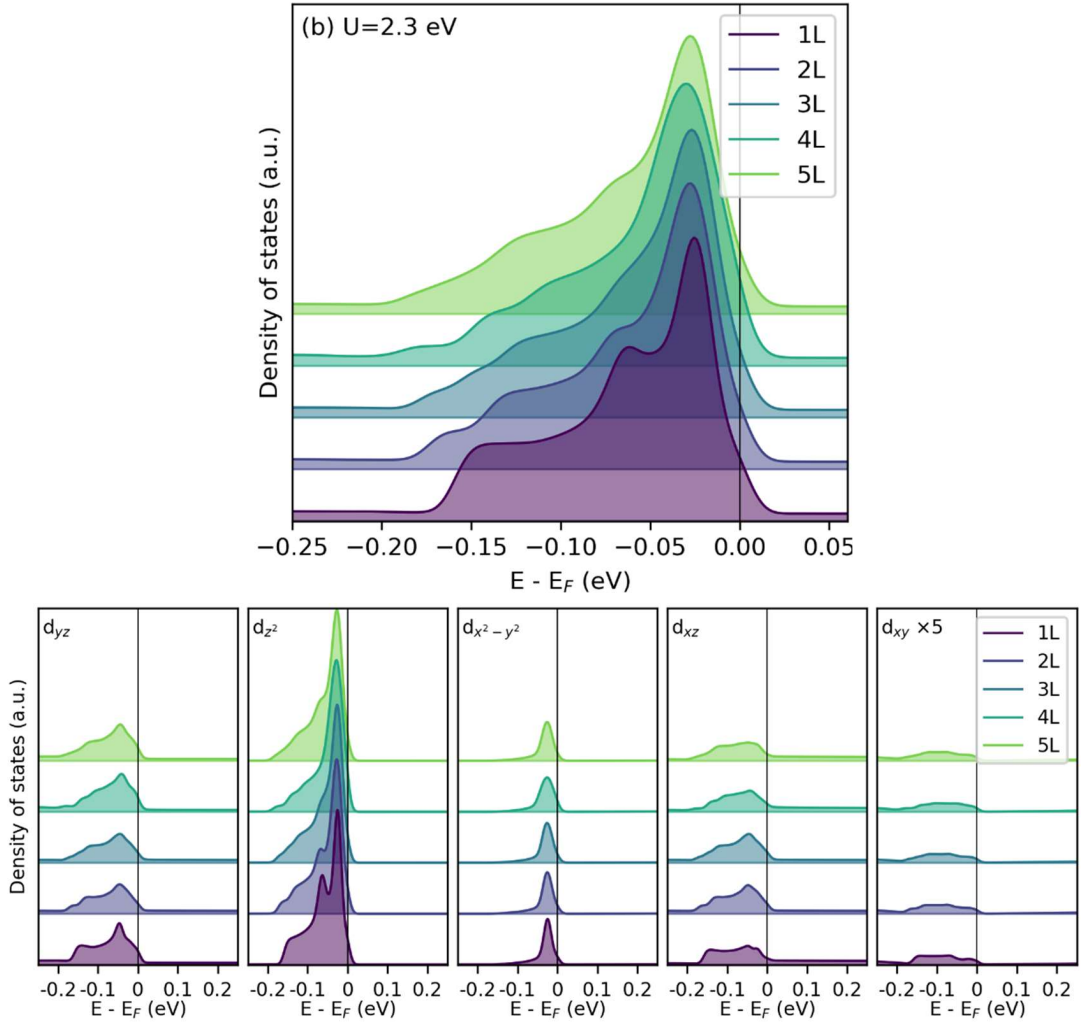

**Fig. 5. Top:** normalized (per layer) total DOS in the case of different layered systems. **Bottom:** the same for different d-orbitals.

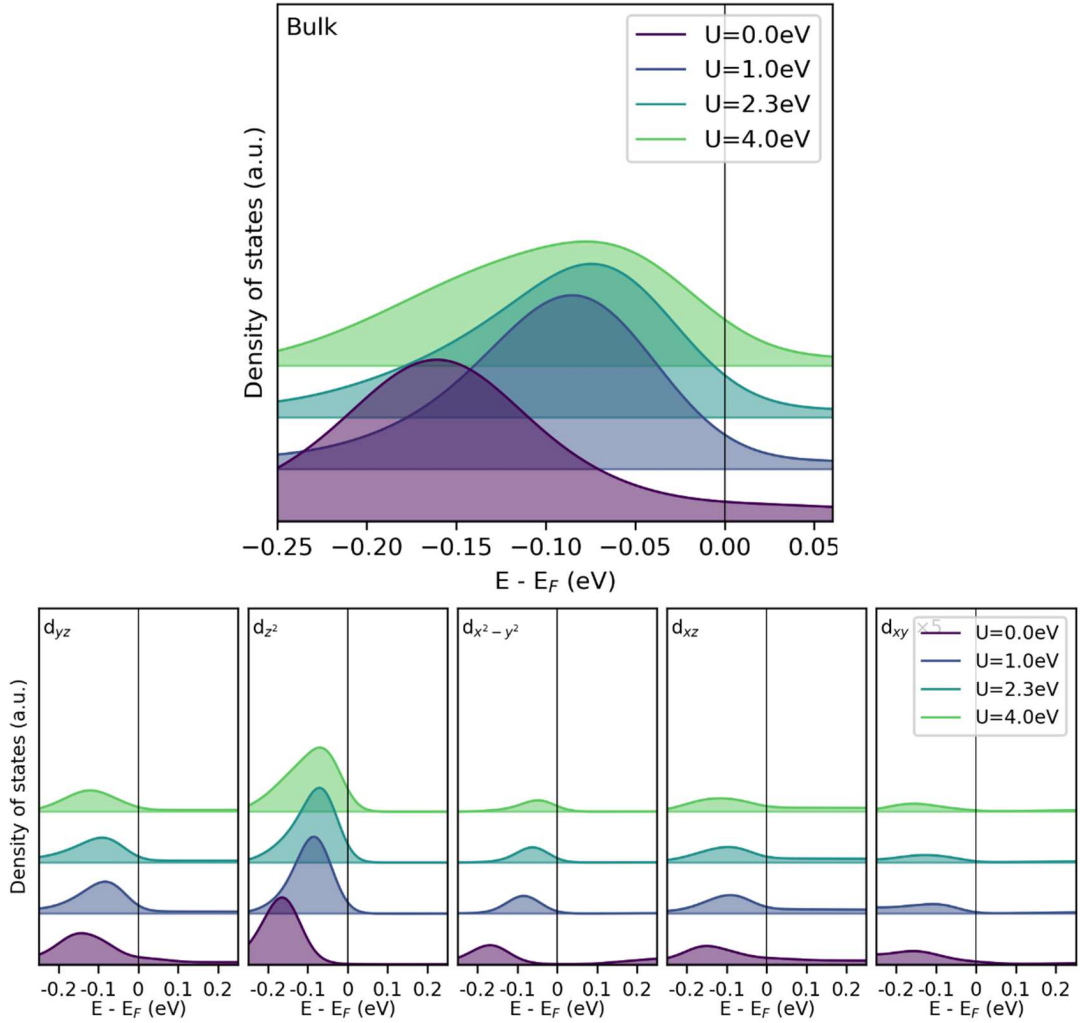

**Figure 6. Top:** normalized total DOS in the case of bulk FeSe with different U values (U=1.0 eV, 2.3 eV, 4.0 eV). **Bottom:** the same for different d-orbitals.

## 4. Conclusions

Through first-principles calculations of the electronic structure and many-body calculations of the transition temperature for the 1L-5L FeSe films, we find that the main contribution to the superconducting pairing comes from the Fe 3d orbital electrons. The difference in this part of the projected DOS for the different thicknesses of the FeSe film is mainly responsible for the difference $T_c$ in the studied systems.

We find that the critical temperature grows with increasing value of the Coulomb repulsion parameter U. In particular, we find that $T_c$~33 K - 53 K for 1L system which is in reasonable agreement with the experimental data for 1L FeSe on STO (65K). Similarly, for the same values of U the results for the bulk system (~ 9.4 K - 21K) is also in a reasonable agreement with experiments (~8K). These results suggests that the rather simple model used in this work may catch the main effects of electron correlation (which concerns mostly the FeSe subsystem, i.e. beyond the substrate). Moreover, we demonstrate in the above that antiferromagnetic fluctuations can produce rather high superconducting $T_c$'s in FeSe films, even without the inclusion of contributions (phonons, doping) from the substrate.

In summary, in this work for the first time we explored scenario of the spin-wave exchange mechanism of superconductivity for FeSe systems with different number of layers. As a test of the approach, we applied it to the bulk FeSe and found a rather good agreement of the critical temperature with experimental data. Thus, we expect a similar agreement in the case of 2L-5L systems, which are expected to be much less affected by the substrate (doping and phonons) than the 1L system. Regarding the 1L FeSe, we show that independently of the substrate spin-wave fluctuations can give a significant contribution to the superconducting critical temperature. Although the used model is rather simple, it helps obtain a meaningful **estimation of the difference of the role of correlations** in the superconductivity in FeSe, as function of layer thickness - a task that has so far not been accomplished with desired accuracy by more refined models. No doubt the accuracy of the obtained results have to be further tested by comparing them with future experimental data, for both isolated systems and systems on substrates.

**Acknowledgements**
The work of the UCF group was supported partially by DOE grant DE-FG02-07ER46354, while that of JS was supported by the China Scholarship Council (CSC) and of GQ by National Natural Science Foundation of China (NSFC) grant No. 11874083.

**Data availability statement**

All data that support the findings of this study are included within the article. Others relevant or raw data are available from the authors upon reasonable request.

61  Zhang, Y. *et al.* Superconducting Gap Anisotropy in Monolayer FeSe Thin Film. *Phys. Rev. Lett.* **117**, 117001, doi:10.1103/PhysRevLett.117.117001 (2016).

62  Subedi, A., Zhang, L., Singh, D. J. & Du, M. H. Density functional study of FeS, FeSe, and FeTe: Electronic structure, magnetism, phonons, and superconductivity. *Phys. Rev. B* **78**, 134514, doi:10.1103/PhysRevB.78.134514 (2008).

63  Kresse, G. & Furthmüller, J. Efficient iterative schemes for ab initio total-energy calculations using a plane-wave basis set. *Phys. Rev. B* **54**, 11169-11186, doi:10.1103/PhysRevB.54.11169 (1996).

64  Blöchl, P. E. Projector augmented-wave method. *Phys. Rev. B* **50**, 17953-17979, doi:10.1103/PhysRevB.50.17953 (1994).

65  Perdew, J. P., Burke, K. & Ernzerhof, M. Generalized Gradient Approximation Made Simple. *Phys. Rev. Lett.* **77**, 3865-3868, doi:10.1103/PhysRevLett.77.3865 (1996).

66  Sigrist, M. Introduction to Unconventional Superconductivity. *AIP Conference Proceedings* **789**, 165-243, doi:10.1063/1.2080350 (2005).

67  Kontani, H., Tazai, R., Yamakawa, Y. & Onari, S. Unconventional density waves and superconductivities in Fe-based superconductors and other strongly correlated electron systems. *arXiv preprint arXiv:2209.00539* (2022).

68  Bazhirov, T. & Cohen, M. L. Effects of charge doping and constrained magnetization on the electronic structure of an FeSe monolayer. *J. Condens. Matter Phys.* **25**, 105506 (2013).

69  Bardeen, J., Cooper, L. N. & Schrieffer, J. R. Theory of Superconductivity. *Phys. Rev.* **108**, 1175-1204, doi:10.1103/PhysRev.108.1175 (1957).

70  Mandal, S., Zhang, P., Ismail-Beigi, S. & Haule, K. How Correlated is the FeSe/SrTiO$_3$ System? *Phys. Rev. Lett.* **119**, 067004, doi:10.1103/PhysRevLett.119.067004 (2017).

# Supplementary Information
# Thickness dependence of superconductivity of FeSe films

Jia Shi,[1,2] Duy Le,[2] Volodymyr Turkowski,[2] Naseem Ud Din,[2] Tao Jiang,[2] Qiang Gu,[1]* Talat S. Rahman[2]*

[1]University of Science and Technology Beijing, Physics, Beijing 100083, China

[2] Department of Physics, University of Central Florida, FL32816, USA

## 1. DFT+U results for the electronic structure of films

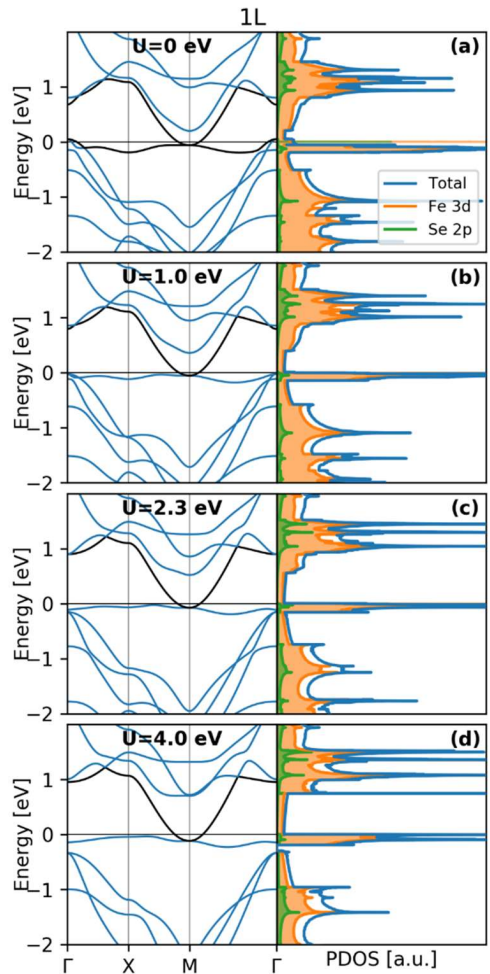

**Figure S1**: (Color online) Band structure and density of states of 1L FeSe film at different values

of U. The horizontal black line marks the Fermi level.

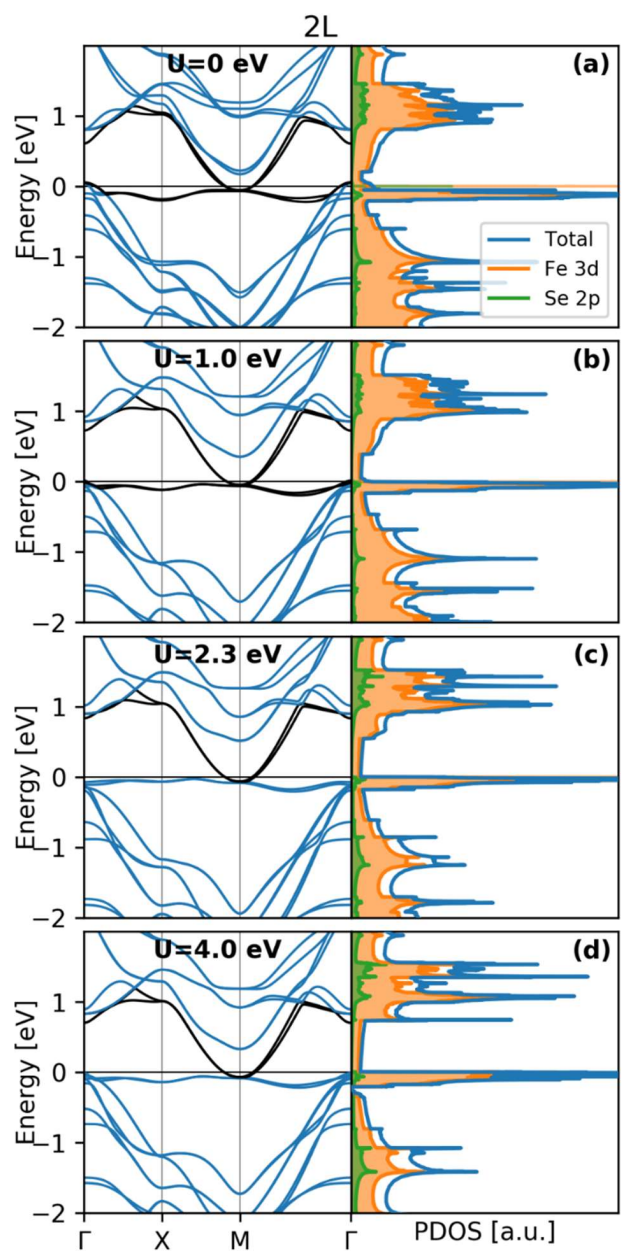

**Figure S2**: (Color online) The same as in Fig. S1 for the 2L system.

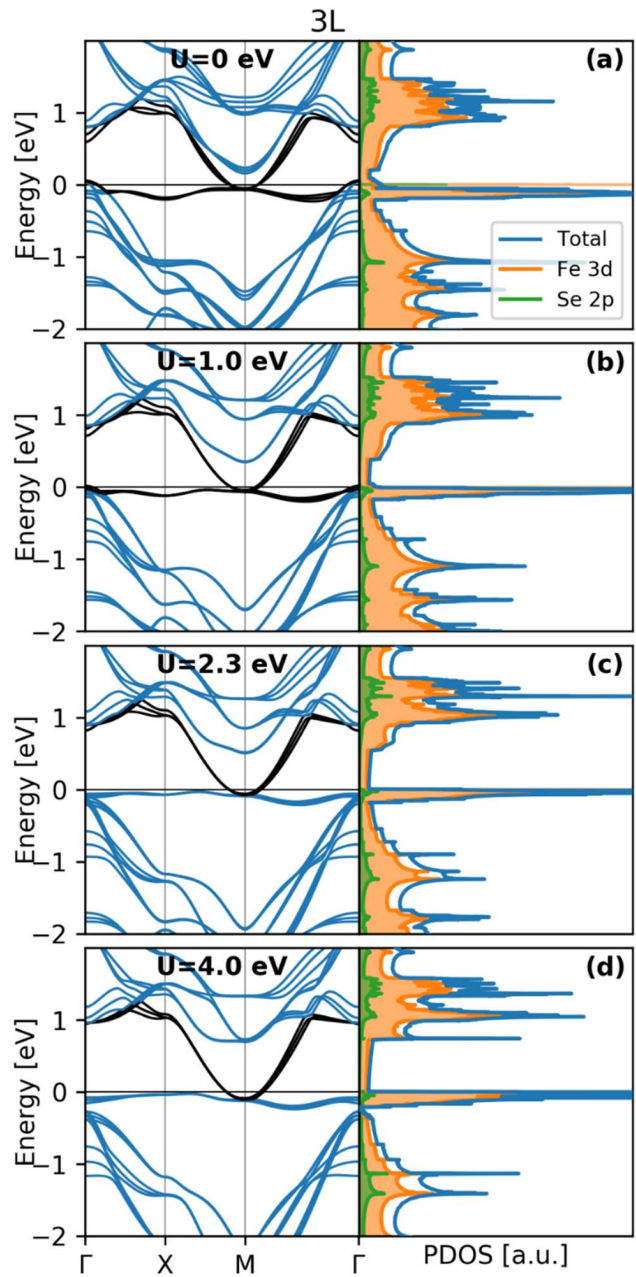

**Figure S3**: (Color online) The same as in Fig. S1 for the 3L system.

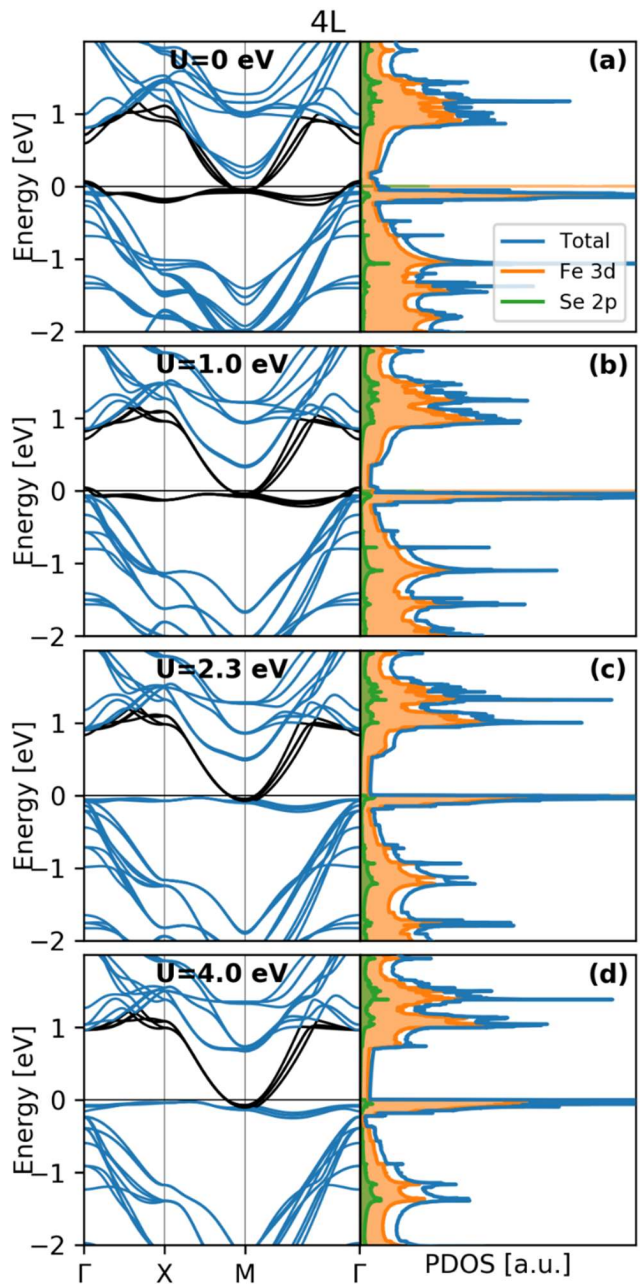

**Figure S4**: (Color online) The same as in Fig. S1 for the 4L system.

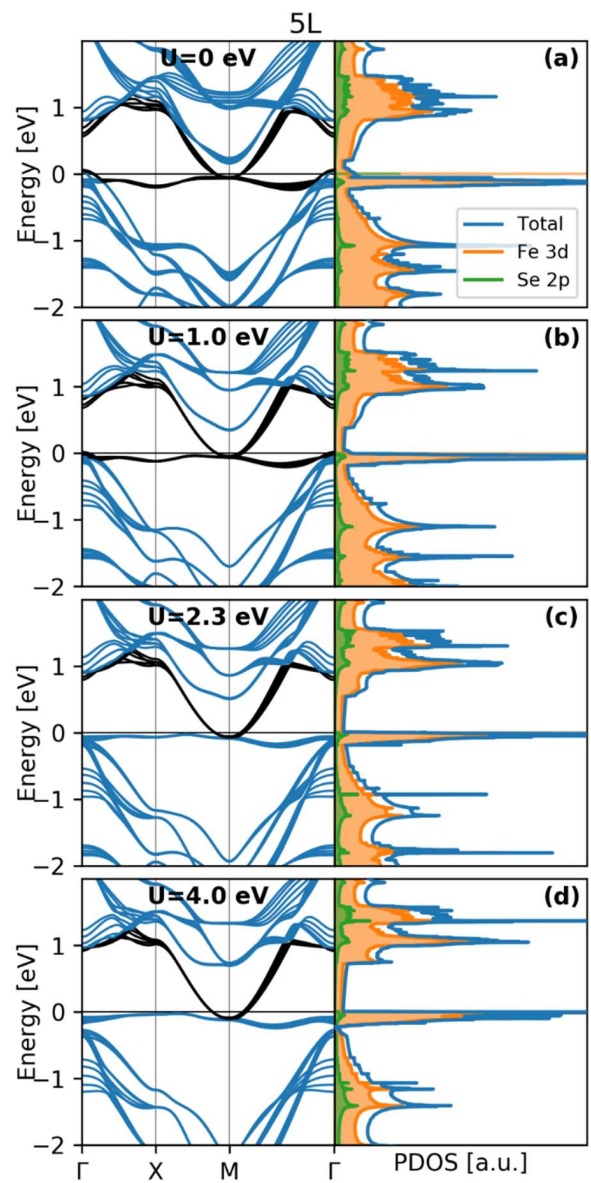

**Figure S5**: (Color online) The same as in Fig. S1 for the 5L system.

## 2. DFT and DFT+U results for the Fermi surface of films

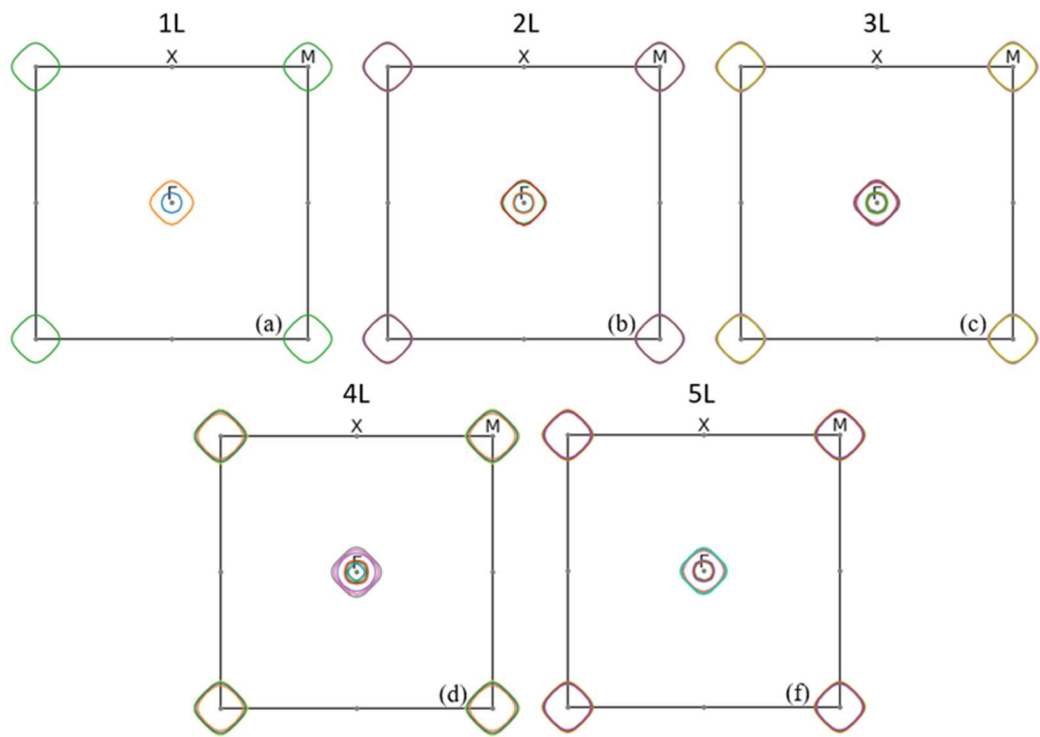

**Figure S6**: (Color online) Fermi surfaces of FeSe films obtained with DFT.

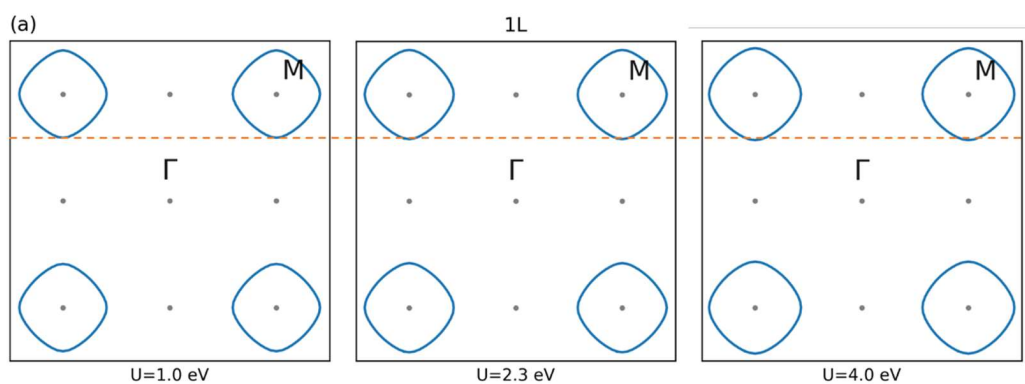

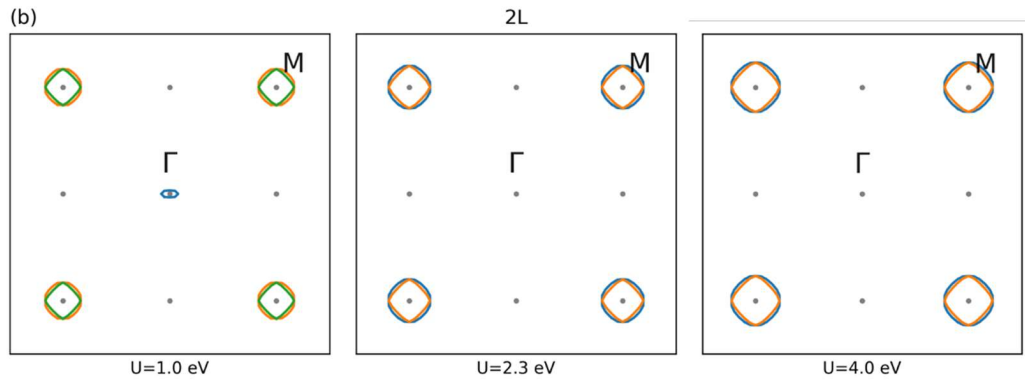

(b) 2L — U=1.0 eV, U=2.3 eV, U=4.0 eV

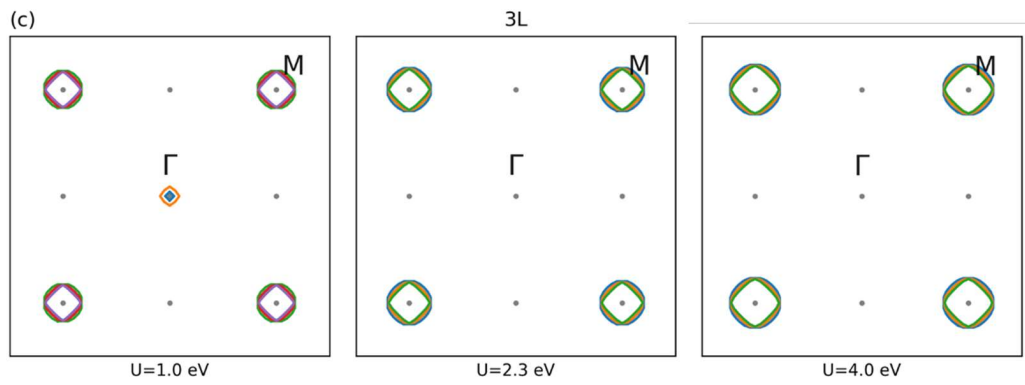

(c) 3L — U=1.0 eV, U=2.3 eV, U=4.0 eV

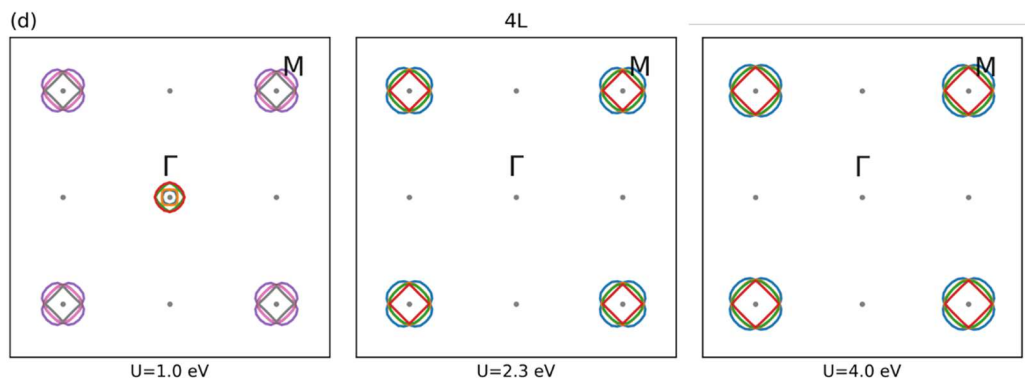

(d) 4L — U=1.0 eV, U=2.3 eV, U=4.0 eV

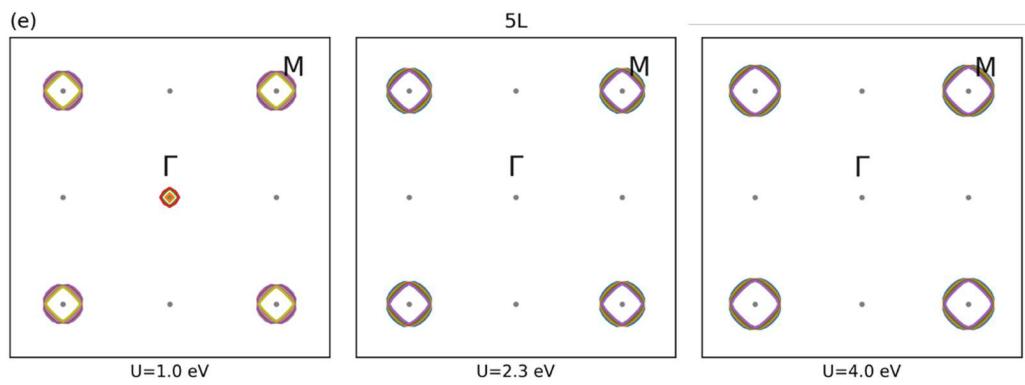

(e) 5L — U=1.0 eV, U=2.3 eV, U=4.0 eV

**Figure S8**: (Color online) Fermi surfaces of FeSe films obtained with DFT+U at different values of U.

## 3. DFT and DFT+U results for the band structure of bulk FeSe

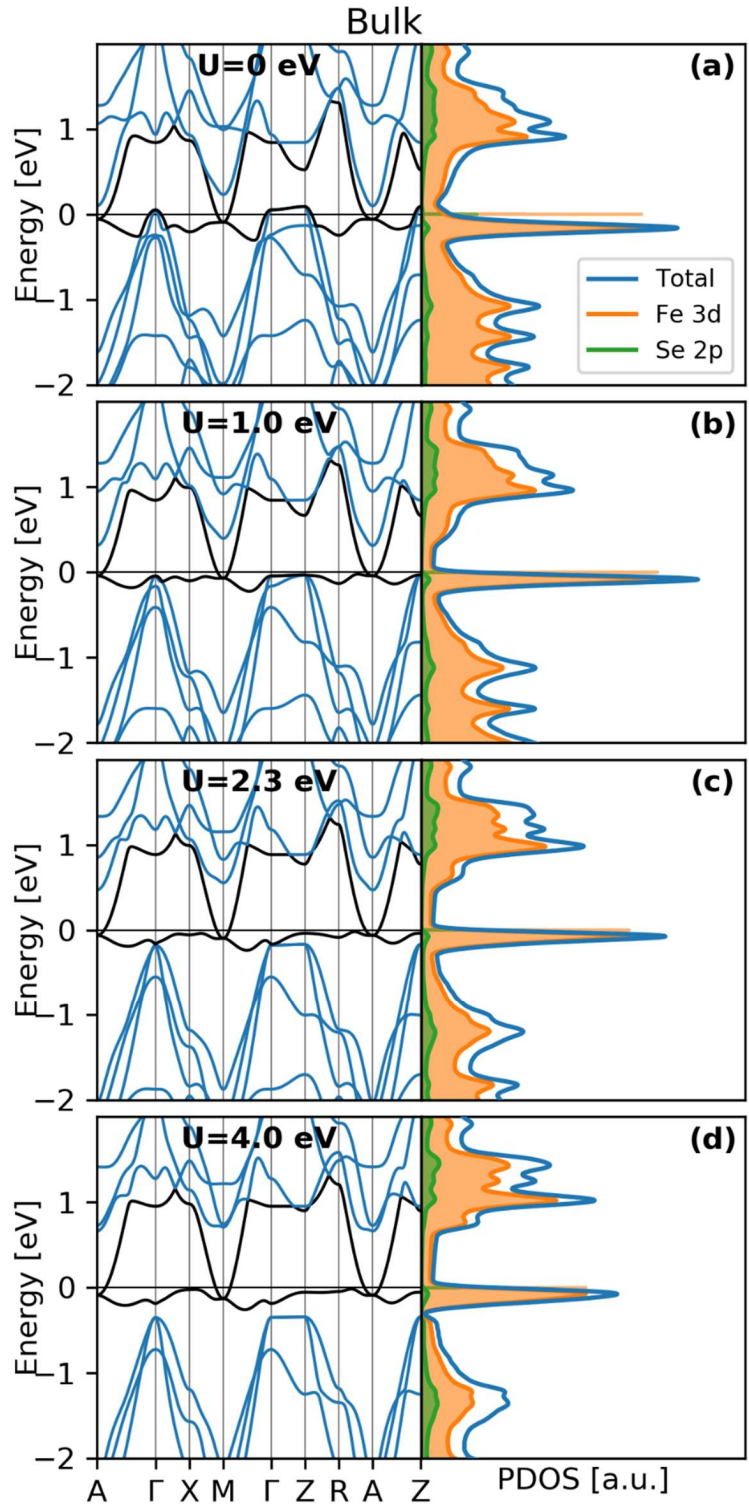

**Figure S9**: (Color online) Polarized band structure of bulk FeSe obtained with DFT (a) and DFT+U at U=1.0 eV (b), 2.3 eV (c), 4.0 eV (d).

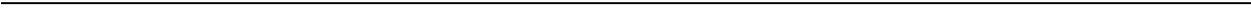

## 4. DFT and DFT+U results for the Fermi surface of bulk FeSe

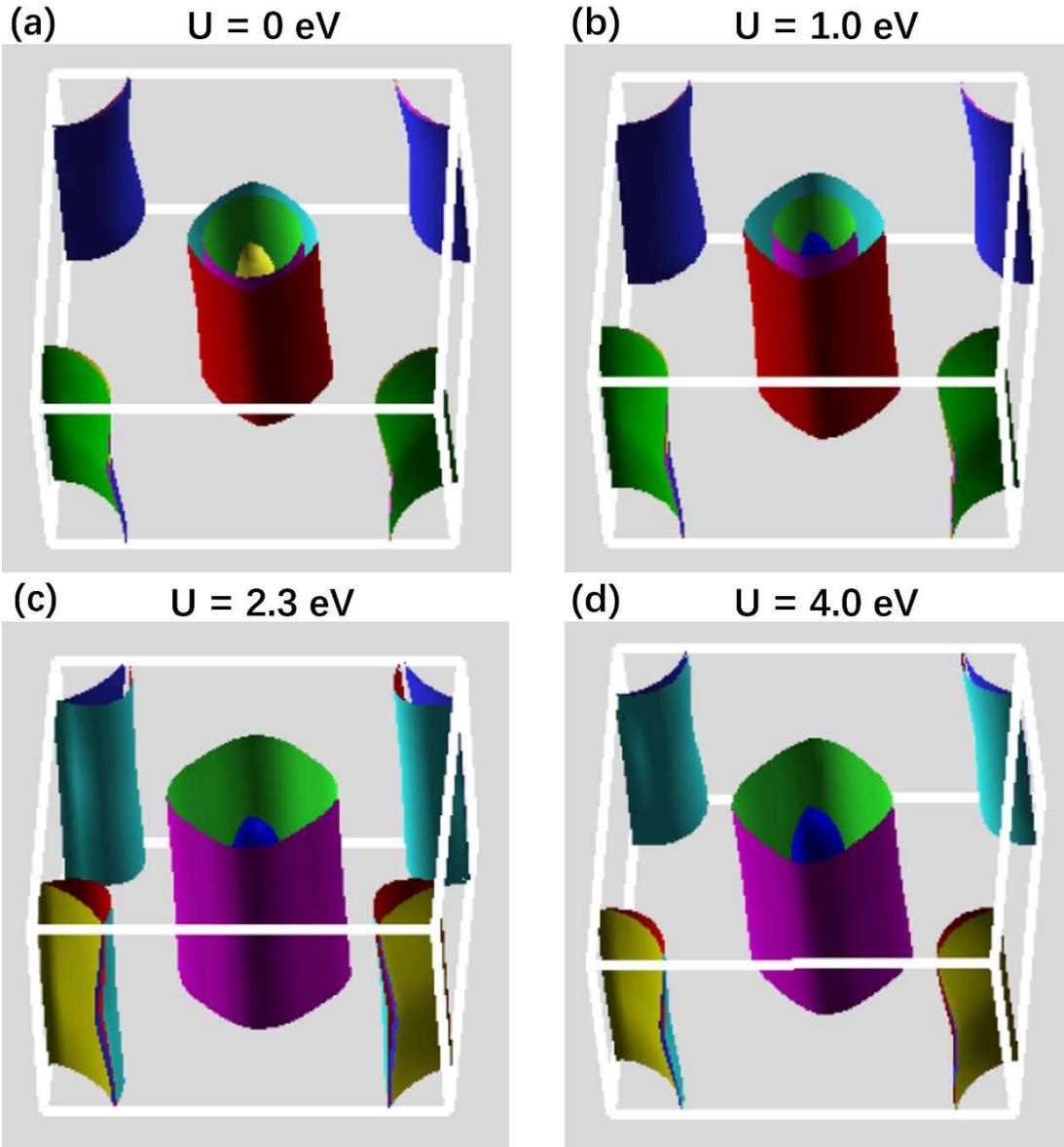

**Figure S10**: (Color online) Spin non-polarized Fermi surfaces of bulk FeSe obtained with DFT (a) and DFT+U at U=1.0 eV (b), 2.3 eV (c), 4.0 eV (d).

# 5. <u>**DFT+U results for the normalized density of states of FeSe systems**</u>

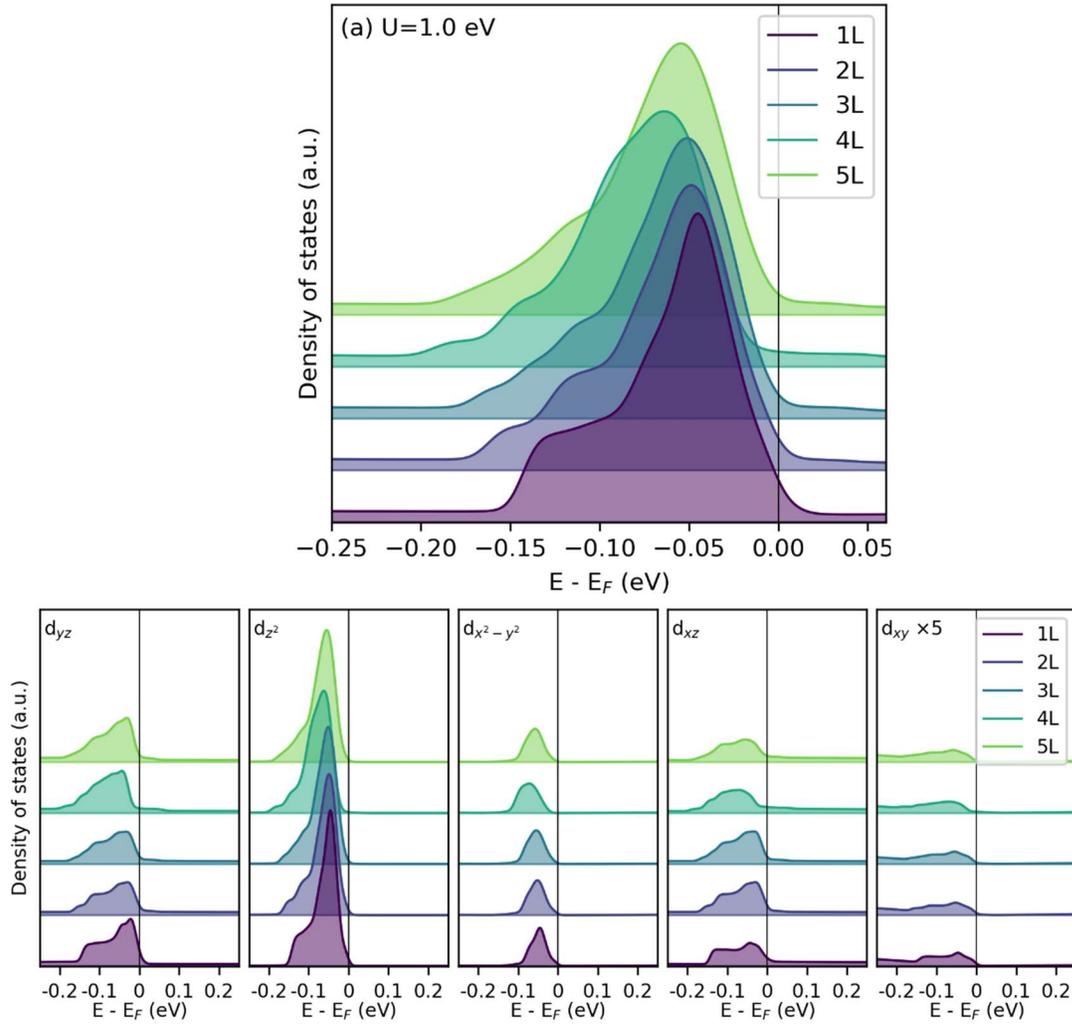

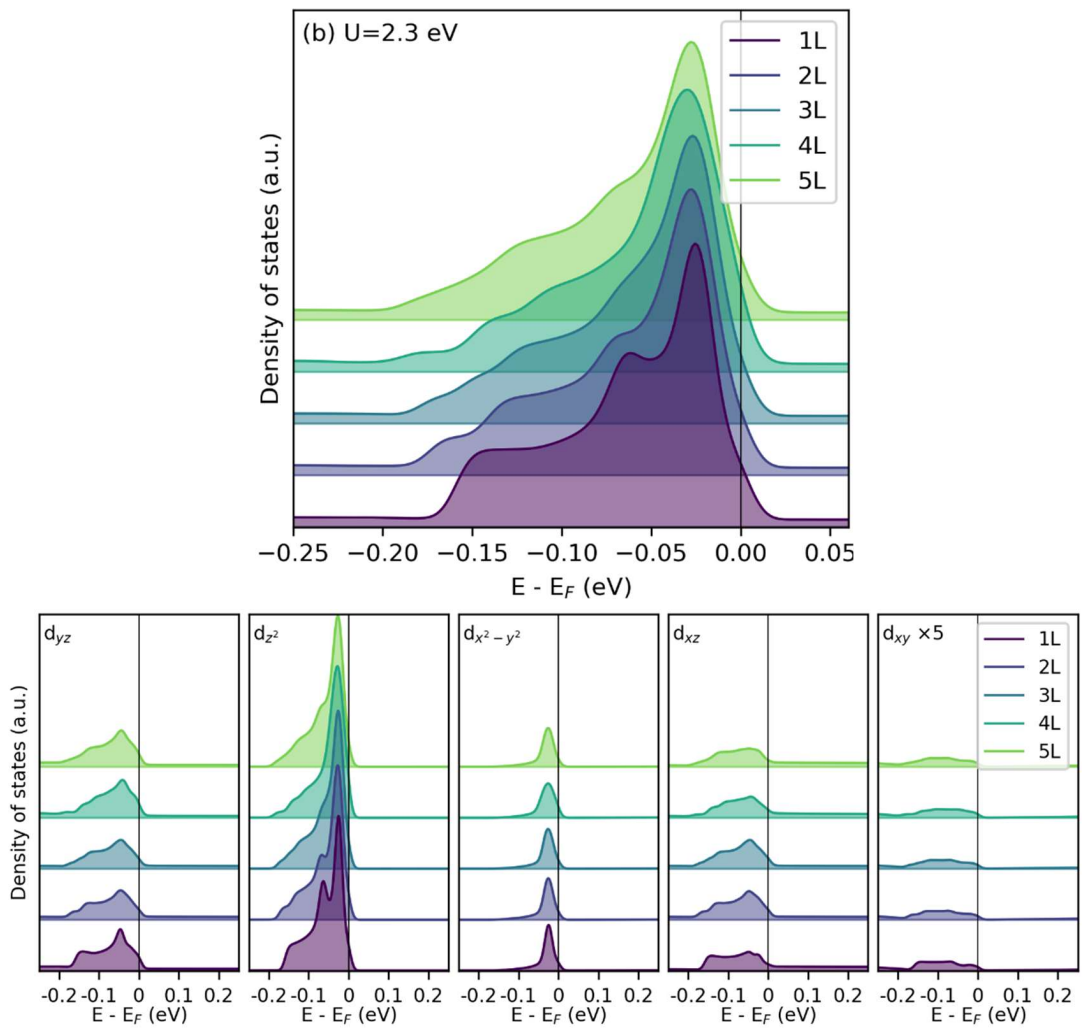

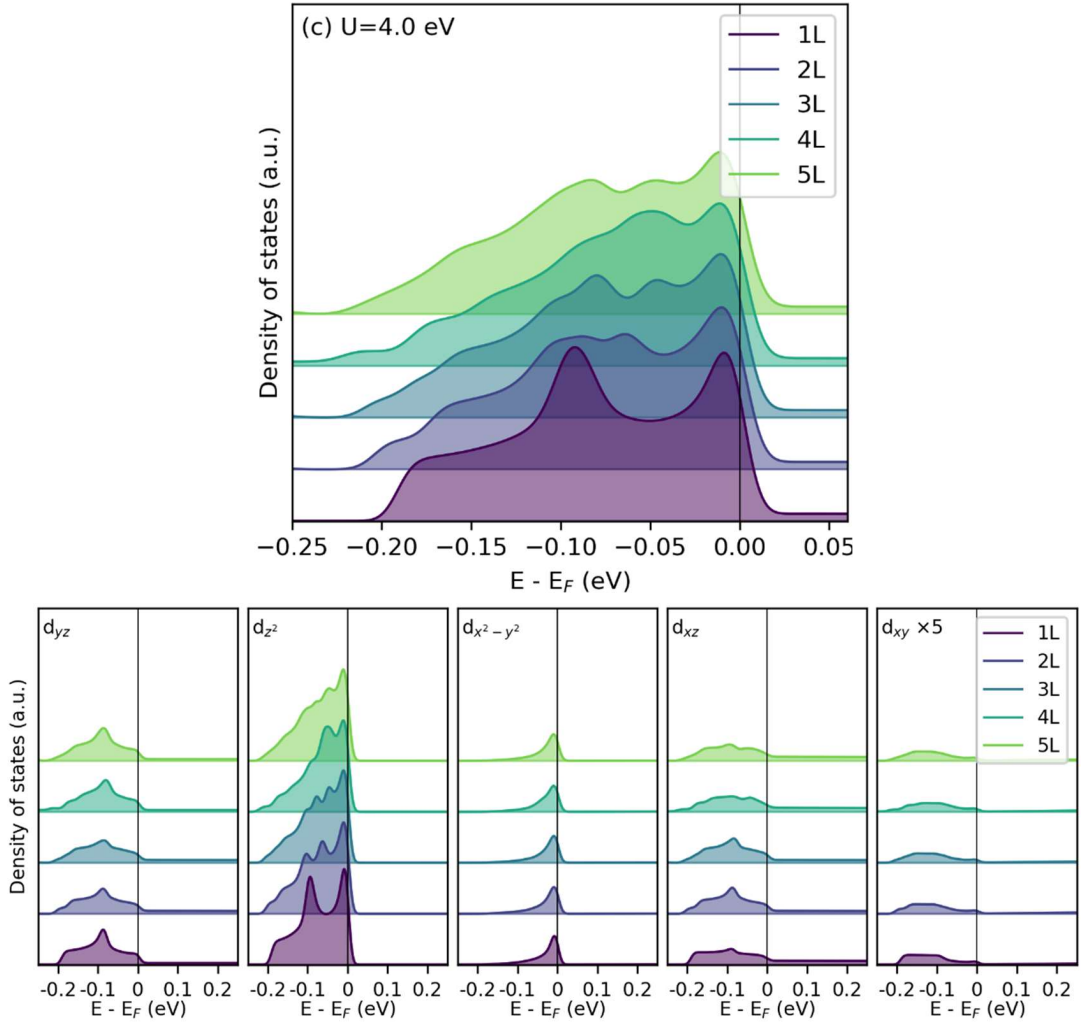

**Figure S11. Top:** normalized (per layer) total DOS in the case of 1L-5L FeSe films. **Bottom:** the same for the d-orbital resolved case. The results are shown for different values of U (U=1.0 eV (a), 2.3 eV (b), 4.0 eV (c)).

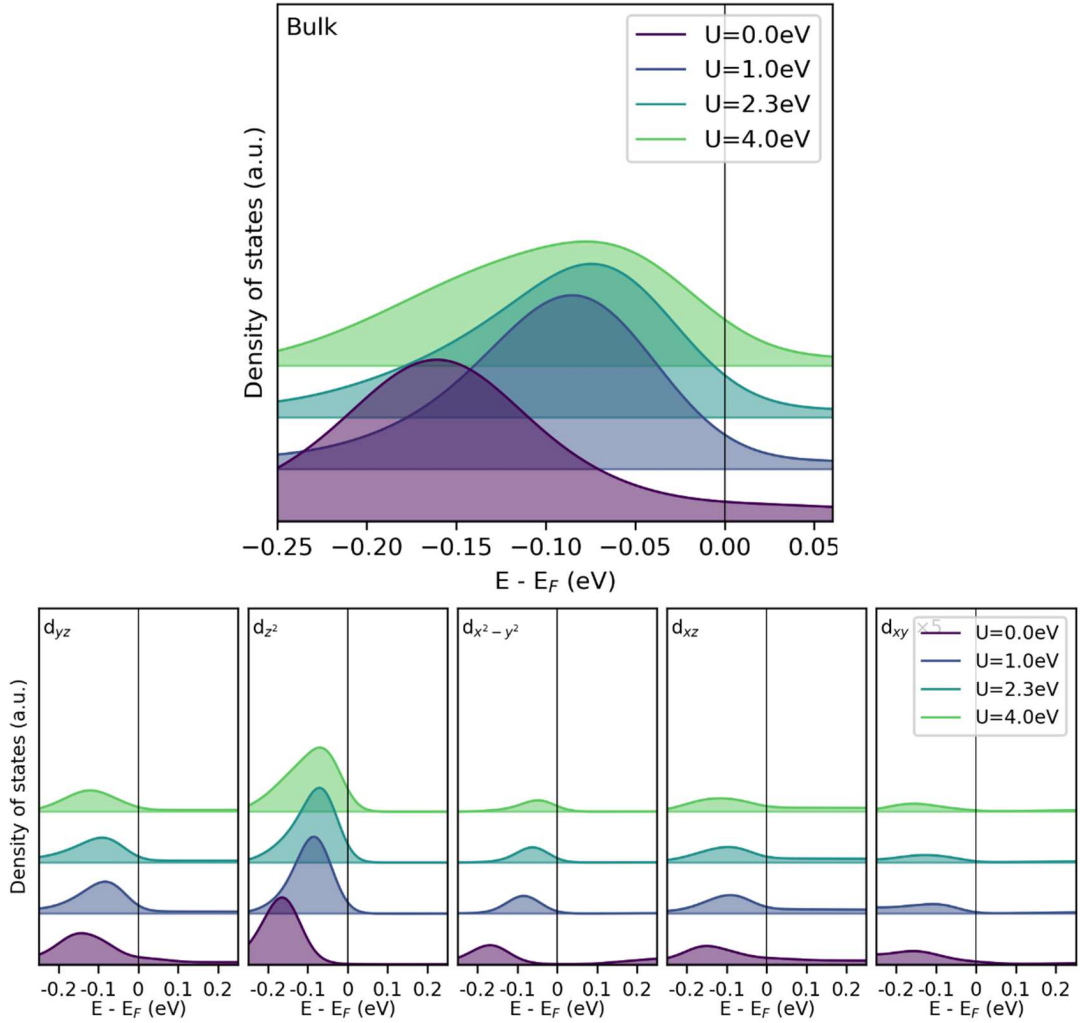

**Figure S12. Upper:** Normalized total DOS in the case of bulk FeSe with different U values (U=1.0 eV, 2.3 eV, 4.0 eV). **Below:** the same for different d-orbitals.

# 6. DFT and DFT+U results for the (antiferromagnetic) magnetic moment

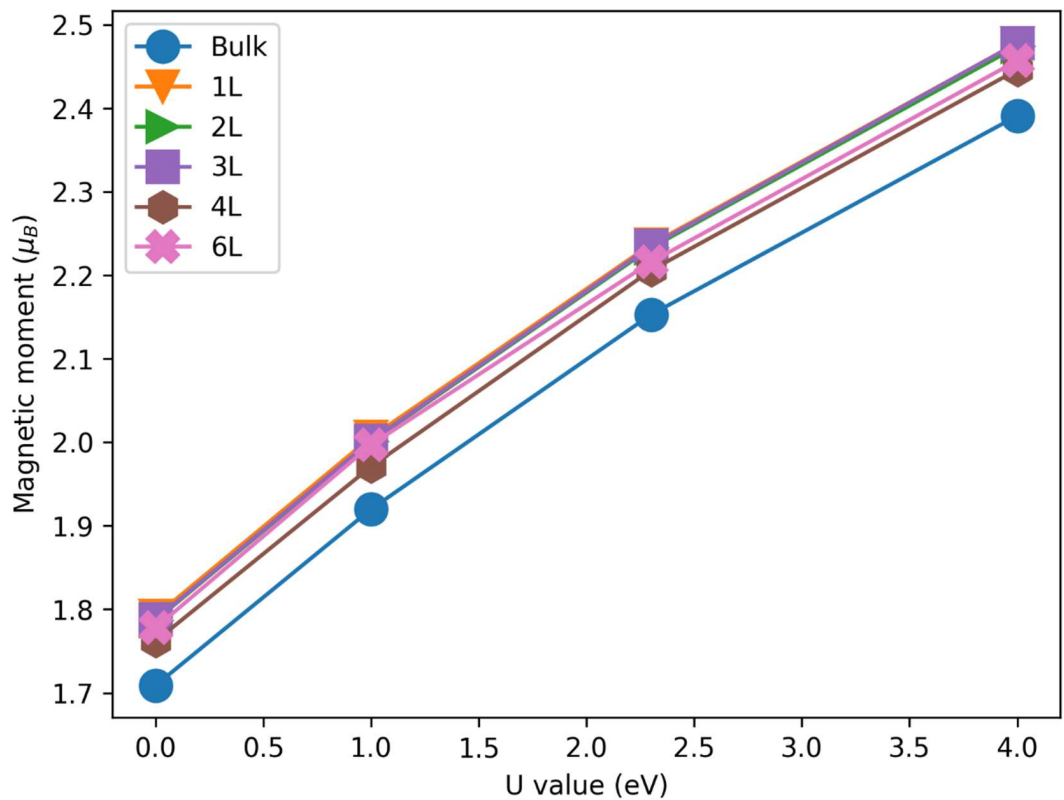

**Figure S13**: (Color online) Antiferromagnetic magnetic moment of 1L-5L FeSe films at different values of U.

## 7. <u>DFT+U results for susceptibility for FeSe films</u>

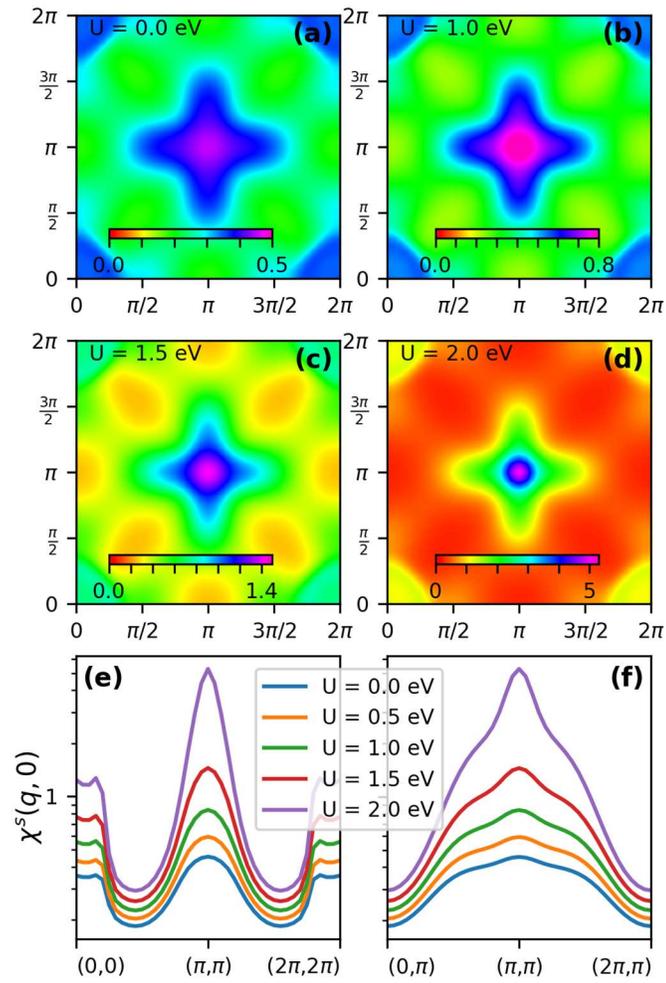

**Figure S14**: (Color online) (a)-(d) Spin susceptibility of 1L FeSe at different values of U. (e)-(f) The same along different momentum paths.

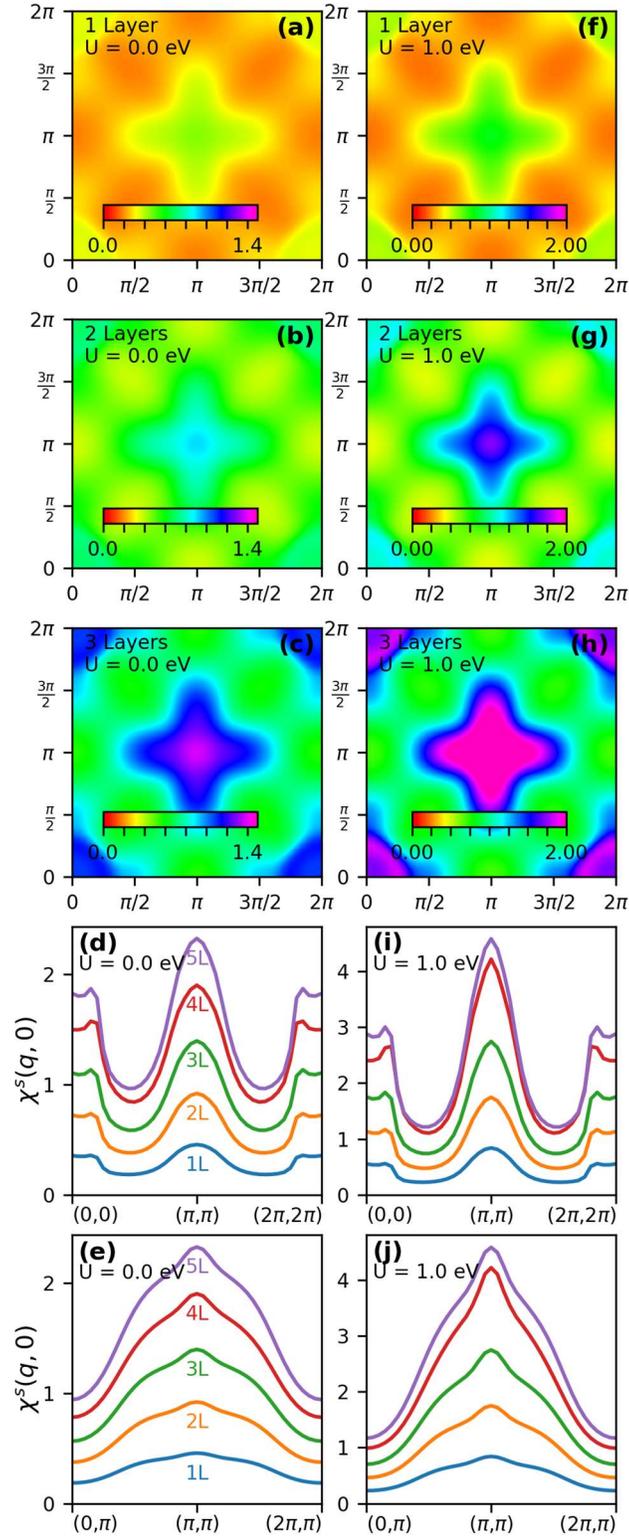

**Figure S15**: (Color online) The same as in Fig. S14 for systems with different numbers of layers.